\documentclass[12pt]{article}
\usepackage{epsf,epsfig}
\usepackage{graphicx}
\usepackage{amsmath}
\usepackage{amssymb}
\usepackage[cp1251]{inputenc}

\title{\large \bf Dark energy as a cosmological consequence of existence\\ of the Dirac scalar field} 
\author{O.V. Babourova\footnote{E-mail: baburova@orc.ru},
B.N. Frolov\footnote{E-mail: frolovbn@orc.ru}\\
{\em Moscow Pedagogical State  University,}\vspace{-2.5mm}\\
{\em Faculty of Physics and Informational Technologies,}\vspace{-2.5mm}\\
{\em M. Pirogovskaya ul. 29, Moscow 119992, Russian Federation}}\vspace{-2.5mm} 
\date{}

\begin{document}
\maketitle
\begin{abstract}
The solution of the field equations of the conformal theory of gravitation with Dirac scalar field in Cartan--Weyl spacetime at the very early Universe is obtained. In this theory dark energy (describing by an effective cosmological constant) is a function of the Dirac scalar field $\beta$. This solution describes the exponential decreasing of $\beta$ at the inflation stage and has a limit to a constant value of the dark energy at large time. This can give a way to solving the fundamental cosmological constant problem as a consequence of the fields dynamics in the early Universe.
\end{abstract}
Pacs 04.50.Kd, 04.20.Fy, 98.80.Jk

\section{\large Introduction}

The Poincar\'{e}--Weyl gauge theory of gravitation (PWTG) has been developed in \cite{BFZ1}. This theory is invariant both concerning the Poincare subgroup and the Weyl subgroup  --  extensions and compressions (dilatations) of spacetime. Dilatations are equivalent in the mathematical sence to the transformations of the group of length calibres changes, which is the gauge group of the H. Weyl theory developed in 1918 \cite{Weyl}. The gauge field introduced by the subgroup of dilatations is named as dilatation field. 
Its vector-potential is the Weyl vector, and its strength is the Weyl's segmental curvature tensor arising in the geometrical interpretation of the theory together with the curvature and torsion tensors. The dilatation gauge field does not coinside with electromagnetic field (that has been asserted by Weyl in his basic work \cite{Weyl}), but represents a field of another type \cite{Ut1}. In particular, quanta of this field can have nonzero rest masses.

As it has been shown in \cite{BFZ1}, an additional scalar field $\beta(x)$ is introduced in PWTG  as an essential geometrical addendum to the metric tensor. The properties of this field coincide with those of the scalar field introduced by Dirac in his well-known article \cite{Dir} and also by S. Deser in \cite{Deser}. We shall name this field as Dirac scalar field. The Dirac scalar field plays an important role in construction of the gravitation Lagrangian, some members of which have structure of the Higgs Lagrangian and after spontaneous braking of dilatational invariance can cause an appearance of nonzero rest masses of particles \cite{Fr_Singap}.  

On the basis of the observational data, it is accepted in modern cosmology that the dark energy (described by the cosmological constant) is of dominant importance in dynamics of the universe. In this connection the major unsolved problem of modern fundamental physics is very large difference of around 120 orders of magnitude between a very small value of Einstein cosmological constant $\Lambda$, which can be estimated on the basis of modern observations in cosmology, and the value of the cosmological constant in the early Universe, which has been estimated by theoretical calculations in quantum field theory of quantum fluctuation contributions to the vacuum energy \cite{Wein}--\cite{China}. In the present work we try to understand the cosmological constant problem as the effect of the gravitational field and the Dirac scalar field dynamics in the Cartan--Weyl spacetime in the early Universe.

\section{\large Gravitational Lagrangian}

Spacetime in PWTG has the geometrical structure of the Cartan--Weyl space with a curvature 2-form  ${\cal R}^a{}_{b}$, a torsion 2-form ${\cal T}^a$ and a nonmetricity 1-form ${\cal Q}_{ab}$ of the Weyl type, ${\cal Q}_{ab} =\frac14 g_{ab}{\cal Q}$, where $\cal Q$ is a  Weyl 1-form. 

On the basis of PWTG, a conformal theory of gravitation in Cartan--Weyl spacetime with the Dirac scalar field has been developed in the external form formalism with the Lagrangian density 4-form $\mathcal{L}$ \cite{BFLip1}--\cite{BabFr_mon},
\begin{equation} 
\mathcal L = \mathcal L_G + \mathcal L_m + \beta ^{4} \Lambda^{ab} \wedge \left(\mathcal Q_{ab} -\frac{1}{4} g_{ab} \mathcal Q \right)\,, \label{eq:fullL}
\end{equation} 
where $\mathcal L_m$ is the matter Lagrangian density 4-form, $\Lambda^{ab}$ are the Lagrange multipliers and 
the gravitational field Lagrangian density 4-form $\mathcal L_G$ reads, 
\begin{equation} 
\begin{split}
\mathcal L_G &=2f_{0} \left[ \frac{1}{2} \beta ^{2} \mathcal R^{a}{} _{b} \wedge \eta _{a}{} ^{b} +\beta ^{4} \Lambda \eta +\frac{1}{4} \lambda \mathcal R^{a}{} _{a} \wedge \ast\mathcal R^{b}{} _{b} +\right. \\ 
& + \tau _{1} \mathcal R ^{[a}{} _{b]} \wedge \ast\mathcal R^{b}{} _{a} +\tau _{2} (\mathcal R^{[ab]} \wedge \theta _{a} )\wedge \ast(\mathcal R^{[c}{} _{b]} \wedge \theta _{c} )+\\
& + \tau _{3} (\mathcal R^{[ab]} \wedge \theta _{c} )\wedge \ast(\mathcal R^{[c}{} _{b]} \wedge \theta _{a} )+\tau _{4} (\mathcal R^{[a}{} _{b]} \wedge \theta _{a} \wedge \theta ^{b} )\wedge \ast(\mathcal R^{[c}{}_{d]} \wedge \theta _{c} \wedge \theta ^{d} )+\\
& + \tau _{5} (\mathcal R^{[a}{} _{b]} \wedge \theta _{a} \wedge \theta ^{d} )\wedge \ast(\mathcal R^{[c}{} _{d]} \wedge \theta _{c} \wedge \theta ^{b} )+ \\ 
& + \tau _{6} (\mathcal R^{a}{} _{b} \wedge \theta _{c} \wedge \theta ^{d} )\wedge \ast(\mathcal R^{c}{} _{d} \wedge \theta _{a} \wedge \theta ^{b} )+ \\
& + \rho _{1} \beta ^{2} \mathcal T^{a} \wedge \ast\mathcal T_{a} +\rho _{2} \beta ^{2} (\mathcal T^{a} \wedge \theta _{b} )\wedge \ast(\mathcal T^{b} \wedge \theta _{a} )+ \\ 
& +\rho _{3} \beta ^{2} (\mathcal T^{a} \wedge \theta _{a} )\wedge \ast(\mathcal T^{b} \wedge \theta _{b} )+\xi \beta ^{2} \mathcal Q\wedge \ast\mathcal Q+\zeta \beta ^{2} \mathcal Q\wedge \theta ^{a} \wedge \ast\mathcal T_{a} + \\
& + l_{1} d\beta\wedge\ast d\beta +l_{2} \beta d\beta \wedge \theta ^{a} \wedge \ast\mathcal T_{a} \left. +l_{3} \beta d\beta\wedge \ast\mathcal Q \frac{}{} \right]+ \\
& + \beta ^{4} \Lambda ^{ab} \wedge (\mathcal Q_{ab} -\frac{1}{4} g_{ab} \mathcal Q).
\label{eq:LG}
\end{split} 
\end{equation}
Here $\wedge$ is the exterior product sign, $\rm d$ is the exterior derivative operator, $*$ is the Hodge dual  conjugation. The second term in (\ref{eq:LG}) is the effective cosmological constant which is interpreted as the dark energy density ($\Lambda$ is the Einstein cosmological constant).

Variational field equations in the Cartan--Weyl spacetime have been derived from $\mathcal{L}$ by exterior form variational formalism \cite{BFKl}. Independent variables are basis 1-forms $\theta^a$, a nonholonomic connection 1-form $\Gamma^a{}_{b}$, the scalar field $\beta$ and Lagrange multipliers $\Lambda^{ab}$. As a result we have $\Gamma$-, $\theta$- and $\beta$-equations, which have the following forms in vacuum ($\mathcal{L}_m\approx 0$) \cite{BFLip1}--\cite{BabFr_mon}, 

$\Gamma$--equation:
\begin{equation} 
\begin{split}
\label{eq:GammaUr1}
2&f_0\left[\beta ^{2}\left(-\frac14\mathcal Q\wedge\eta_{a}{}^{b}
+\frac12\mathcal T^{c}\wedge\eta_{a}{}^{b}{}_{c}+\frac12\eta_{ae}\wedge\mathcal Q^{be}+ d\ln\beta\wedge\eta_{a}{}^{b}\right) + \right. \\
& + \lambda\frac{1}{2} \mathcal D(\delta _{a}^{b} \ast\mathcal R^{c}{} _{c}) + 
\tau_1 2\mathcal D\left(\ast\mathcal R^{[b}{}_{a]}\right)+
\tau_2\mathcal D\left(2\delta _{a}^{[d} \delta _{c]}^{b} \theta _{d}\wedge\ast(\mathcal R^{[fc]} \wedge \theta _{f} )\right) + \\
& + \tau_3\mathcal D\left(2\delta _{a}^{[d}\delta _{c]}^{b}\theta _{f}\wedge\ast(\mathcal R^{[fc]}\wedge\theta _{d})\right) +
\tau_4 2\mathcal D\left(\ast(\mathcal R^{c}{} _{d}\wedge\theta _{c}\wedge\theta ^{d})\theta _{a}\wedge\theta ^{b}\right) + \\
& + \tau_5 2\mathcal D\left(\ast(\mathcal R^{[c}{}_{d]}\wedge\theta _{c}\wedge\theta ^{[b})\theta _{a]}\wedge\theta ^{d}\right) + \tau_6 2\mathcal D\left(\ast(\mathcal R^{c}{} _{d}\wedge\theta _{a} \wedge\theta ^{b})\theta _{c} \wedge \theta ^{d}\right)+\\
& + \rho _{1}2\beta ^{2}\theta ^{b}\wedge\ast\mathcal T_{a} + \rho _{2}2\beta ^{2}\theta ^{b}\wedge\theta _{c}\wedge\ast(\mathcal T^{c} \wedge\theta _{a}) + \rho _{3}2\beta ^{2}\theta ^{b}\wedge\theta _{a}\wedge\ast(\mathcal T^{c} \wedge\theta _{c}) + \\
& + \xi 4\beta ^{2}\delta _{a}^{b}\ast\mathcal Q + \zeta\beta ^{2}\left(2\delta _{a}^{b}\theta ^{c}\wedge \ast\mathcal T_{c} +\theta ^{b}\wedge \ast(\mathcal Q\wedge\theta _{a})\right) + 
l_2\beta\theta ^{b}\wedge\ast(d\beta\wedge\theta _{a}) + \\
& +\left.  l_3 2\beta\delta _{a}^{b}\ast d\beta \frac{}{}\right] - 2\beta ^{4}\Lambda _{a}{} ^{b}=0.
\end{split}
\end{equation}

$\theta$--equation:
\begin{equation} 
\begin{split}
\label{eq:TettaUr1}
2&f_0\left[\beta ^{2}\left(\frac12\mathcal R^{b}{}_{c}\wedge\eta_{b}{}^{c}{}_{a}\right) + 
\beta ^{4}\Lambda\eta _{a} + \right. \\
& + \lambda\left(\frac{1}{4}\mathcal R^{c}{} _{c}\wedge\ast(\mathcal R^{b}{} _{b}\wedge\theta _{a})+\frac{1}{4}\ast(\ast\mathcal R^{b}{} _{b}\wedge\theta _{a})\wedge\ast\mathcal R^{c}{}_{c}\right) + \\
& + \tau_1\left(\mathcal R^{[a}{}_{b]}\wedge\ast\left(
\mathcal R^{b}{}_{a}\wedge\theta_{c}\right)+\ast\left(\ast\mathcal R^{b}{}_{a}\wedge\theta_{c}\right)\wedge\ast\mathcal R^{[a}{}_{b]} \right) + \\
& + \tau_2\left( 2\mathcal R_{[ab]}\wedge\ast(\mathcal R^{[cb]}\wedge\theta _{c} ) -\ast(\mathcal R^{[b}{} _{c]}\wedge\theta _{b}\wedge\theta _{a})\mathcal R^{[dc]} \wedge\theta _{d} - \right. \\
& - \left. \ast(\ast(\mathcal R^{[b}{} _{c]}\wedge\theta _{b})\wedge\theta _{a})\wedge \ast(\mathcal R^{[dc]} \wedge \theta _{d})\right) + \\
& + \tau_3\left( 2\mathcal R^{[b}{} _{c]}\wedge\ast(\mathcal R_{[a}{} ^{c]} \wedge\theta _{b})-\ast(\mathcal R^{[b}{} _{c]}\wedge\theta _{d}\wedge\theta _{a})\wedge \mathcal R^{[dc]}\wedge\theta _{b} - \right. \\
& - \left. \ast(\ast(\mathcal R^{[b}{} _{c]}\wedge\theta _{d})\wedge\theta _{a} )\wedge\ast(\mathcal R^{[dc]}\wedge\theta _{b}) \right) + \\
& + \tau_4\left(\ast(\mathcal R^{[b}{} _{c]}\wedge\theta _{b}\wedge\theta ^{c} )(4\mathcal R_{[af]}\wedge\theta ^{f} + 
\ast(\mathcal R^{[e}{} _{f]}\wedge\theta _{e}\wedge\theta ^{f})\eta _{a}) \right) + \\
& + \tau_5\left(\ast(\mathcal R^{[b}{} _{c]}\wedge\theta _{b}\wedge\theta ^{d})(2\mathcal R_{[ad]}\wedge\theta ^{c} -
2\delta _{a}^{c}\mathcal R_{[fd]}\wedge\theta ^{f} + \right.\\
&+\left. \ast(\mathcal R^{[f}{} _{d]}\wedge\theta _{f}\wedge\theta ^{c})\eta _{a})\right) + \\
& + \tau_6\left( \ast(\mathcal R^{[b}{} _{c]}\wedge\theta _{e}\wedge\theta ^{f})(\ast(\mathcal R^{e}{} _{f}\wedge\theta _{b}\wedge\theta ^{c})\eta _{a} + 2g_{ab} \mathcal R^{e}{} _{f}\wedge\theta ^{c}-\right.\\
&\left. - 2\delta _{a}^{c}\mathcal R^{e}{} _{f}\wedge\theta _{b}\frac{}{})\right)+ \\
& + \rho _{1}\beta ^{2}\left(\frac{}{}2\mathcal D(\ast\mathcal T_{a}) + \mathcal T^{c}\wedge\ast (\mathcal T_{c}\wedge\theta _{a}) + \ast(\ast\mathcal T_{c}\wedge\theta _{a})\wedge\ast\mathcal T^{c} +\right.\\
&\left.+ 4d\ln\beta\wedge\ast\mathcal T_{a} \frac{}{}\right) + \\
& + \rho _{2}\beta ^{2}\left(\frac{}{}2\mathcal T^{d}\wedge\ast(\mathcal T_{a}\wedge\theta _{d}) + 
2\mathcal D(\theta _{b}\wedge\ast(\mathcal T^{b}\wedge \theta _{a})) +\right.\\
&+\left. 4d\ln\beta\wedge\theta _{b}\wedge\ast(\mathcal T^{b}\wedge \theta _{a}) - \right.\\
&  - \left.\ast(\ast(\mathcal T ^{c}\wedge\theta _{d})\wedge\theta _{a})\wedge\ast(\mathcal T^{d}\wedge \theta _{c})- \ast(\mathcal T^{b}\wedge\theta _{c}\wedge\theta _{a})(\mathcal T^{c}\wedge\theta _{b} ) \right ) + \\
& + \rho _{3}\beta ^{2}\left(\frac{}{}2\mathcal D(\theta _{a}\wedge\ast(\mathcal T^{b}\wedge\theta _{b}))+ 
 2\mathcal T_{a}\wedge\ast(\mathcal T^{b}\wedge\theta _{b})-\right.\\
&-\left.\ast(\mathcal T^{b}\wedge\theta _{b}\wedge\theta _{a})(\mathcal T^{c}\wedge\theta _{c}) 
 - \ast(\ast(\mathcal T^{b}\wedge \theta _{b})\wedge\theta _{a})\wedge \ast(\mathcal T^{c}\wedge\theta _{c}) +\right.\\ 
&+\left. 4d\ln\beta\wedge\theta _{a}\wedge\ast(\mathcal T^{b}\wedge\theta _{b})\right ) + \\
& + \xi\beta ^{2}\left(\frac{}{}-\mathcal Q\wedge\ast(\mathcal Q\wedge\theta _{a} )-\ast(\ast\mathcal Q\wedge\theta _{a})\ast\mathcal Q\right) +\\
&+\zeta\beta ^{2}\left(\frac{}{}\mathcal D\ast(\mathcal Q\wedge\theta _{a})-\mathcal Q\wedge\ast\mathcal T_{a}+ 
 \mathcal Q\wedge\theta ^{b}\wedge\ast(\mathcal T_{b}\wedge\theta _{a})+\right. \\ 
& +\left. 2d\ln\beta\wedge\ast(\mathcal Q\wedge\theta _{a}) + \ast(\ast\mathcal T_{b}\wedge\theta _{a})\wedge\ast(\mathcal Q\wedge \theta ^{b})\frac{}{}\right) +\\
& + l_1\left(\frac{}{}-d\beta\wedge\ast(d\beta\wedge\theta _{a}) - \ast(\ast d\beta\wedge\theta _{a})\wedge\ast d\beta\frac{}{}\right) + \\
& +l_2\left(\frac{}{}\beta (\mathcal D\ast(d\beta\wedge\theta _{a}) + 
+ d\beta\wedge\theta ^{b}\wedge\ast(\mathcal T_{b}\wedge\theta _{a}) - 
d\beta\wedge\ast\mathcal T_{a} + \right. \\
&+ \left. \ast(\ast\mathcal T_{b}\wedge\theta _{a})\wedge\ast(d\beta \wedge\theta ^{b})) + \right.\\ 
& + \left.\left. d\beta\wedge\ast(d\beta\wedge\theta _{a})\frac{}{}\right) + 
l_3\beta\left(\frac{}{} -d\beta\wedge\ast(\mathcal Q\wedge\theta_{a})-\ast(\ast\mathcal Q\wedge\theta _{a})\ast d\beta\frac{}{}\right)\right]=0. 
\end{split}
\end{equation}

$\beta$--equation:
\begin{equation} 
\begin{split}
\label{eq:BettaUr}
2&f_{0}  \left[ \frac{}{}\beta \mathcal R^{a}{} _{b} \wedge \eta _{a}{} ^{b} -4\beta ^{3} \Lambda \eta +2\rho _{1} \beta \mathcal T^{a} \wedge \ast\mathcal T_{a} + \right. \\ 
& + 2\rho _{2} \beta (\mathcal T^{a} \wedge \theta _{b} )\wedge \ast(\mathcal T^{b} \wedge \theta _{a} )+2\rho _{3} \beta (\mathcal T^{a} \wedge \theta _{a} )\wedge \ast(\mathcal T^{b} \wedge \theta _{b} ) + \\ 
& + 2\xi \beta \mathcal Q\wedge \ast\mathcal Q + 2\zeta \beta \mathcal Q \wedge \theta ^{a} \wedge \ast \mathcal T_{a} +l_{1} (-2d\ast d\beta )+l_{2} (-\beta d(\theta ^{a} \wedge \ast \mathcal T_{a} ))+ \\
& \left. + l_{3} (-\beta d\ast\mathcal Q)\frac{}{}\right] +4\beta ^{3} \Lambda ^{ab} \wedge (\mathcal Q_{ab} -\frac{1}{4} g_{ab} \mathcal Q)=0.
\end{split}
\end{equation}

The variation with respect to the Lagrange multipliers $\Lambda^{ab}$ gives the Weyl condition for the nonmetricity 1-form ${\cal Q}_{ab}$,
\begin{equation} 
\begin{split}
\label{eq:LambdaUr1}
\mathcal Q_{ab} - \frac{1}{4}g_{ab} \mathcal Q=0.
\end{split}
\end{equation}

\section{\large Solutions of the field equations at ultra-early Universe}

We shall solve the field equations for the scale factor $a(t)$ and the scalar Dirac field $\beta$ at the very early stage of evolution of universe, when a matter density has been very small, $\mathcal{L}_m\approx 0$. We shall omit the terms quadratic in curvature  for symplicity.

In  homogeneous and isotropic spacetime the conditions,  $\mathcal {T}^a = \frac{1}{3} \mathcal {T}\wedge\theta^a $ are valid, and we shall find, as the consequence of the field equations, the torsion and nonmetricity in the forms, 
${\mathcal T}_\mu = \chi_T {\rm d}\ln\beta$, ${\mathcal Q}_\mu = \chi_Q {\rm d}\ln\beta$, whete the coefficients $\chi_T$, $\chi_Q$ are expressed by couple constants of the Lagrangian density (\ref{eq:LG}). 

We consider the spatially flat Friedman--Robertson--Walker (FRW) metric 
\begin{equation}
ds^2 = dt^2 - a^2(t)(dx^2 + dy^2 + dz^2)\,.
\label{eq:Frid}
\end{equation}
Taking into account that $\mathcal{L}_m\approx 0$, we obtain from the $\theta$-equation together with the $\beta-$equation the following system of equations \cite{BFLip1}--\cite{BabFr_mon},
\begin{eqnarray}
(0,\,0):& 3\frac{\dot{a}^2}{a^2} + 6\frac{\dot{a}}{a}\frac{\dot{\beta}}{\beta} 
+ 3B_3 \left (\frac{\dot{\beta}}{\beta}\right )^2 = \Lambda{\beta}^2\,, &\label{eq:ur00}\\
(1,\,1):& 2\frac{\ddot{a}}{a} + 2\frac{\ddot{\beta}}{\beta} +4\frac{\dot{a}}{a}\frac{\dot{\beta}}{\beta} + \left (\frac{\dot a}{a}\right )^2 \nonumber\\
&+ (B_2 - 2) \left (\frac{\dot \beta}{\beta}\right )^2 = \Lambda{\beta}^2\,,& \label{eq:ur33}\\
\beta\;\;:& A\left(\frac{\ddot{\beta}}{\beta} + 3\frac{\dot{a}}{a}\frac{\dot{\beta}}{\beta}\right ) 
+ (B - A) \left (\frac{\dot \beta}{\beta}\right )^2 = 0\,,& \label{eq:ur22}
\end{eqnarray}
where the constants $A$, $B$, $B_3 = \frac{1}{3}(2B_1 + B_2)$, $B_1$, $B_2$ are expressed through the parameters of the Lagrangian density (\ref{eq:LG}), the components (2,\,2) and (3,\,3) being equal to the component (1,\,1).

The system of equations (\ref{eq:ur00})--(\ref{eq:ur22}) is inconsistent, because we have three equations for two unknown functions $a(t)$ and $\beta (t)$. Let us put in this system, $B_1 = B_2 = B_3 = 1$, and also $u= \ln a$, $v = \ln \beta$. Then substract Eq. (\ref{eq:ur00}) from Eq. (\ref{eq:ur22}). As a result we obtain the following system of equations, \begin{eqnarray}
&& (\dot u)^2 + 2\dot{u}\dot{v} + (\dot v)^2= \frac{\Lambda}{3}e^{2v}\,\label{eq:ur0}\\
&& \ddot{u} + \ddot{v} - \dot{u}\dot{v} - (\dot v)^2= 0\,, \label{eq:ur3}\\
&& \ddot{v} + 3\dot{u}\dot{v} + \frac{B}{A } (\dot v)^2 = 0\,. \label{eq:ur2}
\end{eqnarray}
Eq. (\ref{eq:ur0}) is equivalent to the equation, 
\begin{equation}
\dot u + \dot{v} = \pm \frac{\lambda}{3} e^{v}\,,\quad \lambda = \sqrt{3\Lambda}\,. \label{eq:dudv}
\end{equation}
It is easy to check that Eq. (\ref{eq:ur3}) is fulfilled identically as a consequence of Eq. (\ref{eq:dudv}). Therefore 
we have only 2 equations (\ref{eq:ur2}) and (\ref{eq:dudv}) for 2 unknown functions $a(t)$, $\beta (t)$, and this system of equations is consistent. In what follows we choose the sign "+" in Eq. (\ref{eq:dudv}). 

Let us find $\dot u$ from  Eq. (\ref{eq:dudv}) and put it in Eq.(\ref{eq:ur2}). We obtain the equation, 
\begin{equation}
\ddot v - \lambda e^{v}\dot{v} + \omega (\dot v)^2 = 0  \,, \quad \omega = \frac{B}{A} - 3 \,. \label{eq:ddv}
\end{equation}
The first integral of this equation is the following, 
\begin{equation}
\dot v = \lambda_0 e^{-\omega v} - \frac{\lambda}{1+\omega}e^{v}\,, \label{eq:bdot}
\end{equation}
where $\lambda_0$ is a constant of integration. 

The system of equation (\ref{eq:dudv}), (\ref{eq:bdot}) have a large variety of integrable solutions parametrized by $\omega$ and $\lambda_0$. Let us obtain the solution for the case $\omega=0$. If we put in Eq. (\ref{eq:bdot}) $\lambda_0 = \lambda$, then this equation reads, $\dot v = \lambda (1-e^{v})$, and we have a solution \cite{BabFr_mon},   
\begin{equation}
\beta (t) = e^{v(t)} = \frac {1}{1-e^{-\lambda (t + t_0)}}\,, \quad 
a(t) = a_0 e^{\frac{\lambda}{3}(t+t_0)} (1-e^{-\lambda (t+t_0)})^{4/3}\,. \label{eq:bsol1}
\end{equation}

We assume that the value of $\beta$ is very large, when $t=0$. Therefore the constant of integration $t_0$ should be very small ($0<t_0\ll \lambda^{-1}$). Then from Eq. (\ref{eq:bsol1}) under $t\gg t_0$ one has approximately, 
\begin{equation}
\beta (t) = (\beta_{01})\,\!^{\exp{(-\lambda t)}}\,, \qquad a(t) = (a_{01})e^{\frac{\lambda}{3}t}\,. \label{eq:bea}
\end{equation}

\begin{figure}[ht]\center
\includegraphics*[height=8cm]{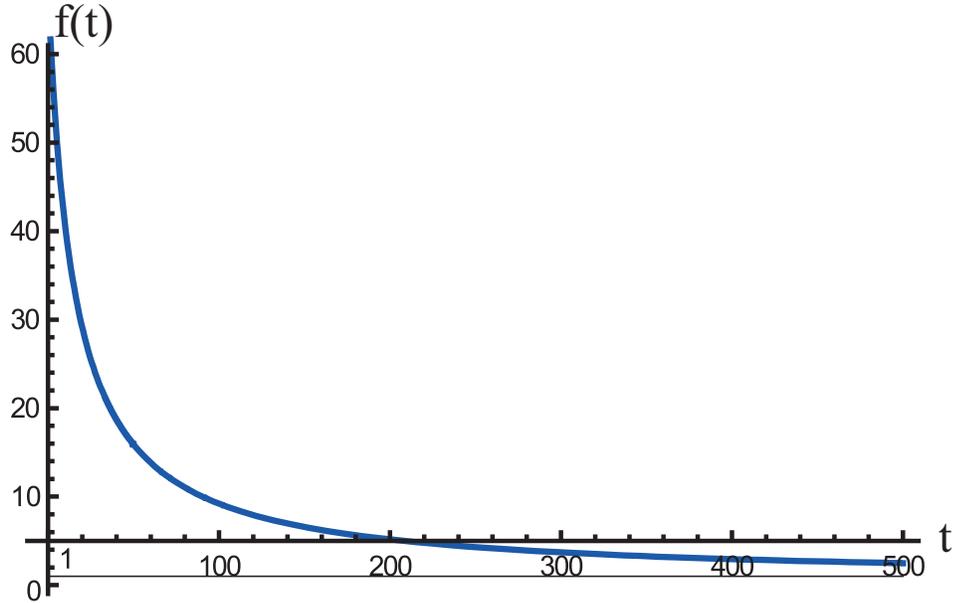}
\caption{The solution (\ref{eq:bsol1}) for the Dirac scalar field in the early Universe}
\label{fig:Ris}
\end{figure}

These solutions realize exponential diminution of a field $ \beta $ 
(see Figure \ref{fig:Ris})  for the function (\ref{eq:bsol1})), and thus sharp exponential decrease of physical vacuum energy (dark energy) by many orders. We have $ \Lambda _ {eff} = \beta^2 \Lambda \rightarrow \Lambda $ in a limit at $t\rightarrow \infty $. Thus, the  effective cosmological constant can slightly differ already by the end of inflation from the limiting value equal to its  modern size $ \Lambda $ that provides the subsequent transition from the Friedman epoch to the epoch of the accelerated expansion in accordance with the modern observant cosmological data.

We have for the solutions (\ref{eq:bsol1}), (\ref{eq:bea}),   
\begin{eqnarray}
&&\beta \rightarrow 1\,, \quad \Lambda_{eff} = \beta^2 \Lambda \rightarrow \Lambda\,, \quad
{\rm when}\quad  t \rightarrow \infty\,.\label{eq:brar}
\end{eqnarray}
Therefore the limit of the effective cosmological constant for large time is not zero (see Figure \ref{fig:Ris}) and is equal to the value of the Einstein cosmological constant that ensures an accelerating expansion of the modern Universe. 

Our solutions are realized, if the following conditions are valid, $B=3A$, $B_1=1$, $B_2=1$. These conditions are determined in rather complicated manner by the 16 coupling constants of the gravitational Lagrangian density (\ref{eq:LG}), and can be easily fulfilled.   

\section{\large Discussion and final remarks}

Here the field equations for the spatially flat homogeneous and isotropic ultra-early Universe are investigated and the solution for the Dirac scalar field (and therefore for the effective cosmological constant) is found. This solution for the scalar field has the behavior of the very extensive diminishing exponent, which limits is the modern value of the Einstein cosmological constant. The solution (\ref{eq:bsol1}) could be realized at the very beginning of the Universe evolution, when the cosmological constant $\Lambda_0$ estimated by quantum field theory was equal $\Lambda_0 /\Lambda = \beta_0^2 \sim 10^{120}$, and the number $\beta_0 \sim 10^{60}$ was very large.  

Thus our result can explain the exponential decrease in time at very early Universe of the dark energy (the energy of physical vacuum), describing by the effective cosmological constant. This can give a way to solving the problem of cosmological constant as a consequence of fields dynamics at the early Universe. It is well-known that this problem is one of the fundamental problems of the modern theoretical physics \cite{Wein}--\cite{China}.

We point out that the ultra-rapid decrease of the energy of physical vacuum according to the law (\ref{eq:bsol1}) occurs only prior to the Friedman era evolution of the Universe. Further evolution of the Universe is determined not by a scalar field only, but also by the born ultra relativistic matter and the radiation interacting with it.

The Dirac scalar field condenses near massive objects. As a consequence of this, the authors formulate the hypothesis that {\it the Dirac scalar field is realized itself not only as the 'dark energy', but also as one of the components of the 'dark matter'} \cite{BabFr_mon}.

\end{document}